# Using Short Message Service (SMS) to Support Business Continuity

Maher Abdel-qader
Department of Computer Science,
Amman Arab University for
Graduate Studies
Amman, Jordan
Maher_abdelqader@hotmail.com

Prof. Dr. Ahmad AL-Jaber
Department of Computer Science,
Amman Arab University for
Graduate Studies
Amman, Jordan
mmmsmkm@yahoo.com

Prof. Dr.Alaa AL-Hamami
Department of Computer Science,
Amman Arab University for
Graduate Studies
Amman, Jordan
Alaa_hamami@yahoo.com

Abstract— Now a day's many organizations are required to communicate online on a daily basis, 24-hour, seven-days-a-week, to gain the desired competitive advantages and profits; although there are a variety of disruptions that may occur within business application such as broken (off-line) database-links and unhanded database exceptions. Such cases will end the automated business work, and force business users to continue business procedures and functionalities via paper work, which causes additional resources with less business competitive advantages. In this paper, we will propose a new model in which we embed short message services (SMS) within business applications using the SMS Gateway such as "Ozeki Message Server", and programmed application packages. By using our proposed model, we can maintain business continuity when a partial disruption occurs and then switch to our application model. As a result to the experimental work, we conclude that our model supports business continuity since it supports the account balance modification while the database link is disrupted. In addition, we carried out each step twice and the scenario was reliable since all of its steps were reliable.

## I. INTRODUCTION

Business and economy automation has been responsible for shifting the world economy from the industrial jobs to the service jobs during the 20th and 21st centuries. Nowadays, business automation is playing a critical role in achieving the required business competitive advantages, in which there are a lot of researches in this field have been performed, most of these researches focus on how to automat a continuing process.

Business functionalities and tasks [1, 2, 4, 5, 6] automation improves the economy of an enterprise, when an enterprise has invested in automation technology to return its investment, or when a state or country increases its income from moving to automation like Germany or Japan in the 20th Century [4, 10].

Business Continuity is the ability to keep vital business operations running in the event of failure. It describes a mentality or methodology of conducting day-to-day business, typically, when a part of the existing infrastructure fails, It is expected to provide a response within a given time period, typically referred to as an SLA (Service Level Agreement). These failures can include power failures, application errors, network failures, data integrity issues, human error or any other issue where the majority of the infrastructure is still in place, but operations are halted and need to resume [3, 8]. The foundation of business continuity are the standards, program development, , supporting policies; guidelines, and procedures needed to ensure a firm to continue without stopping, irrespective of the adverse circumstances or events.

All system design, implementation, support, and maintenance must be based on this foundation in order to have any hope of achieving business continuity, disaster recovery, or in some cases, system support. Business continuity is sometimes misunderstood with disaster recovery, but they are two separated entities, more formally disaster recovery is a small subset of business continuity.

Short Message Service (SMS) is a communication tool that provides a convenient means for people to communicate with each other using text messages via mobile devices or Internet connected computers. Solutions for e-Marketers are available to deliver bulk of SMS messages to large group of people, instead of sending SMS messages one by one manually. Other utilities can collect phone numbers from imported text files or contact information stored in mobile phones [7]. SMS is globally accepted wireless service that enables the transmission of alphanumeric messages between mobile subscribers and external systems such as electronic mail, paging, and voice mail systems.

Ozeki Message Server is a user interface with flexible SMS Gateway application that enables us to send/receive SMS messages to mobile devices with our computer. The application can use a GSM mobile phone attached to the PC with a phone-to-PC data cable or IP SMS technology to transmit and receive messages. Figure 1, illustrates how the Ozeki Message Server interacts with other application components and interactions.

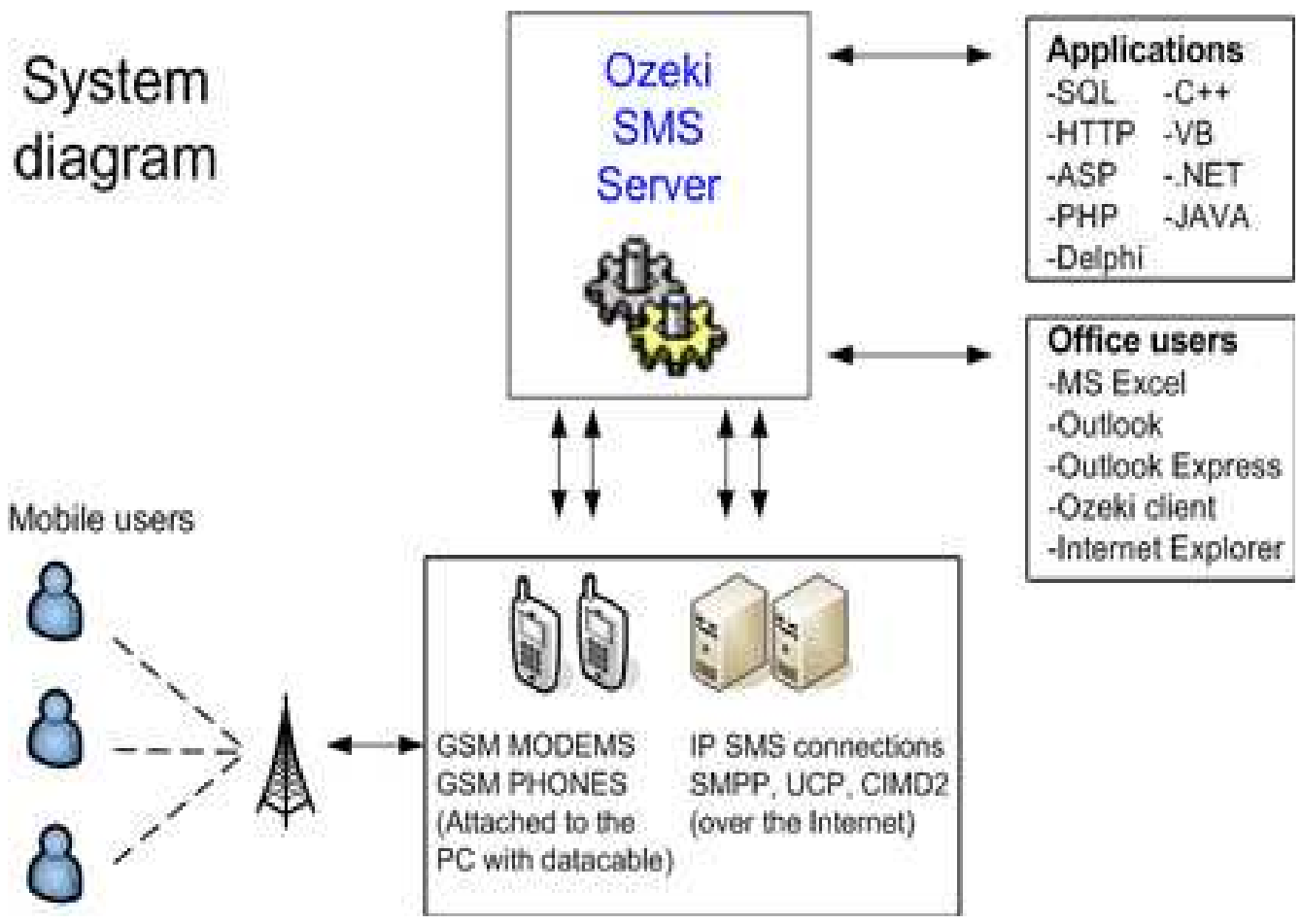

Figure 1. Ozeki Message Server System Diagram [6].

In this paper, we will develop a new model for business continuity, in which we use the SMS technology to maintain the continuity of business operations when a disruption occurs. Our model is about having a standby SMS channel of communication to mitigate opportunities of losing connections among related business sites, this will support business continuity concept while transferring and exchanging critical data and/or business procedures (By coded the database transaction statement and send it via SMS from node to node, and creating package as an interface solution that interact with application to handle any disruption in connection between ( two database nodes), from one database to another when business application connection disruption occurs; in addition, we use SMS also to alert co-coordinators at the right time when failure occurs.

## II. Related Work

There are interests in the field of SMS usages and applications, some of them have proposed a framework that uses SMS as a business tool [5, 11]; in addition, this technology used as an alerting tool in SMS based applications. Furthermore, SMS technology supports remote human/ machine control [12, 13, 14]. In our proposed model, we apply some of the previous mentioned usages of SMS techniques and we deploy SMS technique to support business continuity through developing automated operations that will transmit data from machine to machine and/or from machine to human.

## III. SMS Technology

No secret that wireless technology has become the standard for capacitating communication, entertainment and education across the planet today. In today's organizations, accurate and continuing business procedures highly depend on such technology. One of the most important communication concepts that are based on the wireless technology is the SMS. The sending message (text only) from the mobile is stored in a central short message center (CSMS), and then forwards it to the destination mobile, this means that in case that the recipient is not available; the short message is stored and can be sent later. Each short message can be no longer than 160 characters. These characters can be text (alphanumeric) or binary non-Text short messages. An interesting feature of SMS is the return message from the recipients, which means that the sender, if wishes, can get a small message notifying that the short message has delivered to the intended recipient. Since SMS used signaling channel as opposed to dedicated channels, these messages can be sent/received simultaneously with the voice/data/fax service over a GSM network. The SMS supports national and international roaming. This means that we can send short messages to any other GSM mobile user around the world. With the PCS networks based on all the three technologies, GSM, CDMA and TDMA supporting SMS [7].

Business security is a critical issue that business users should be aware of. In other words, and in terms of SMS security; users should be aware that SMS messages might be subject to interception. Solutions such as encrypted SMS should be considered if there is a need to send sensitive information via SMS [9].

## IV. Business Continuity

Business continuity is the activity performed by an organization to ensure that critical business functions will be available to customers, suppliers, regulators, and other entities that must have access to those functions. These activities include many daily chores such as project management, system backups, change control, and help desk, it is not something implemented at the time of a disaster and moreover it refers to those activities performed daily to maintain service, consistency, and recoverability. Business continuity describes a mentality or methodology of conducting day-to-day business, where it is planning is an activity of determining what this methodology should be. The business continuity plan may be thought as of the incarnation of a methodology that is followed by everyone in an organization on a daily basis to ensure normal operations [8].

The foundation of business continuity are the standards, program development supporting policies; guidelines, and procedures needed to ensure a firm to continue without stoppage, irrespective of the adverse circumstances or events. All system design, implementation, support, and maintenance must be based on this foundation in order to have any hope of achieving business continuity.

## V. Proposed SMS Model to Support Business Continuity

In this research, we build and integrate a business continuity model, in which we use SMS technology and other information system concepts, such as organized databases, SMS listeners, encryption and decryption techniques and SMS

Ozeki Server. Figure 2, illustrates the representative context diagram of our model.

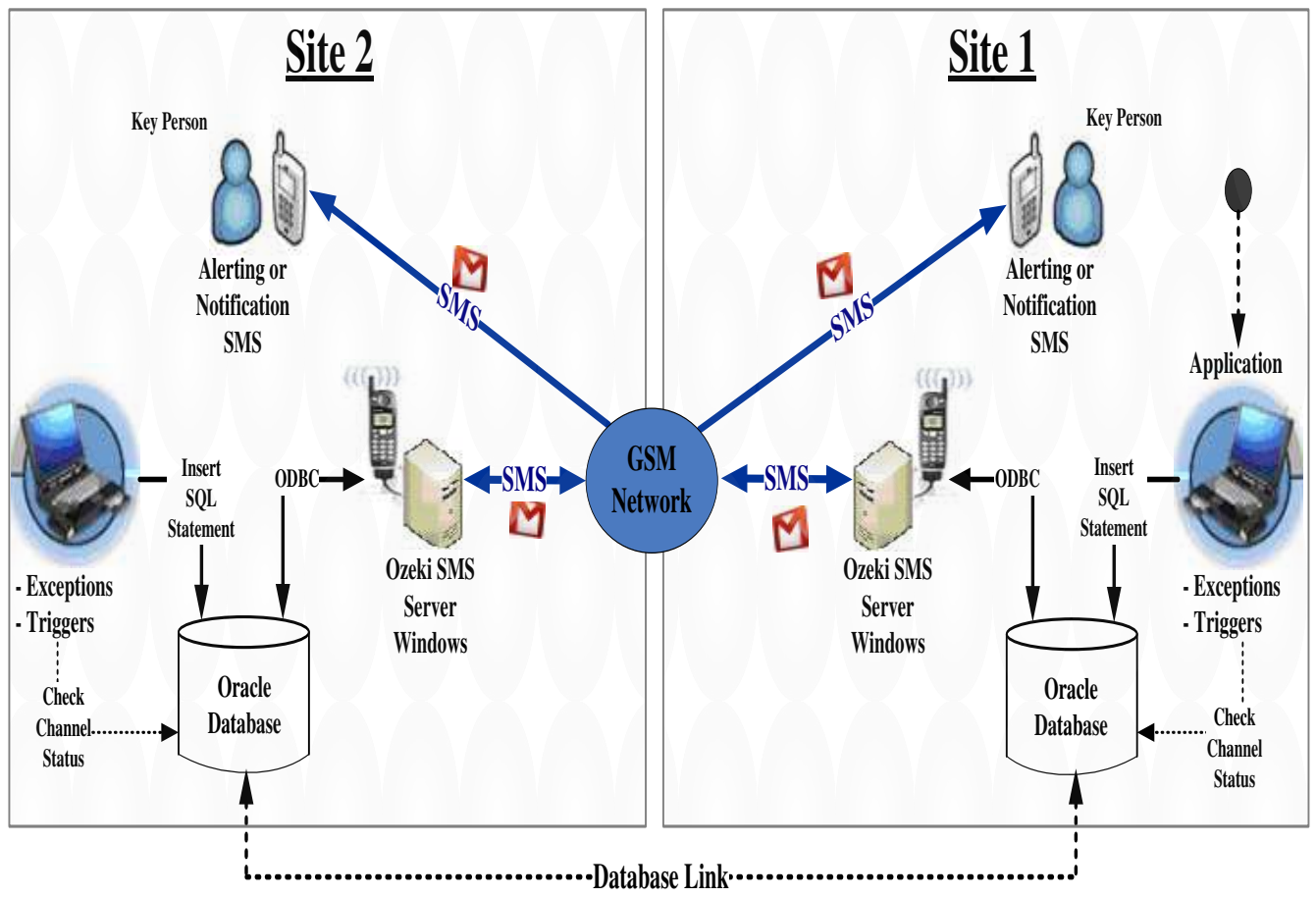

Figure 2. Components and Interactions of our Model

As illustrated in figure 2, two sites are connecting together through a database link, via this link, both of the oracle databases in both sites exchange and execute SQL statements in order to maintain business continuity. In case the database link is broken, there will be no automated business continuity between both sites. Therefore, in our model, we aim to support such case by ensuring business continuity through a standby alternative communication channel.

A. Model's Components and interactions

The starting point of the model can be described as follows:-

when the application in site 1 finds the ability of using the standby SMS channel is allowed and an SMS related trigger or exception exists; in this case, the application is going to insert an SQL statement into oracle database in order to manipulate the uncompleted transaction (exception) or alerting the correspondent parties with a particular suspicious situation (trigger). With regard to SMS exception, it occur when the database link between site 1 and site 2 is broken, and the running SQL transaction has not been completed yet. Now the application is going to insert the two rows as coded formats in the SMS-Log-Table in the oracle database; one for the uncompleted transaction and the other for alerting the key-person. On the other hand, the SMS trigger occurs when a business pre-determined rule exists, for instance, if a banker cashes a check larger than a specified amount with respect to that banker; in this case, the application is going to insert a single row in the SMS-Log-Table in the oracle database in order to alert the key-person with such a situation

B. Model's Application Package

In our model, we deploy our application package for site 1, the package is responsible for ending and starting the database-job and the job is executing the programmed listener; in addition the application package is responsible for encrypting and inserting the message to the SMS_LISTENER_LOG table, finally handling exceptions and triggers in order to insert the corresponding messages into the SMS_LISTENER_LOG table in site 1. Also the package is responsible of making the decryption and fetching from OZEKIMESSAGEIN table in site 2. Finally, execute the incoming messages in the database of site 2. Since the application package exists in both sites (1 and 2), therefore the opposite of previous operations take place if we start from site 2, figure 3 and 4 illustrate application package components and Flowchart of the programmed listener.

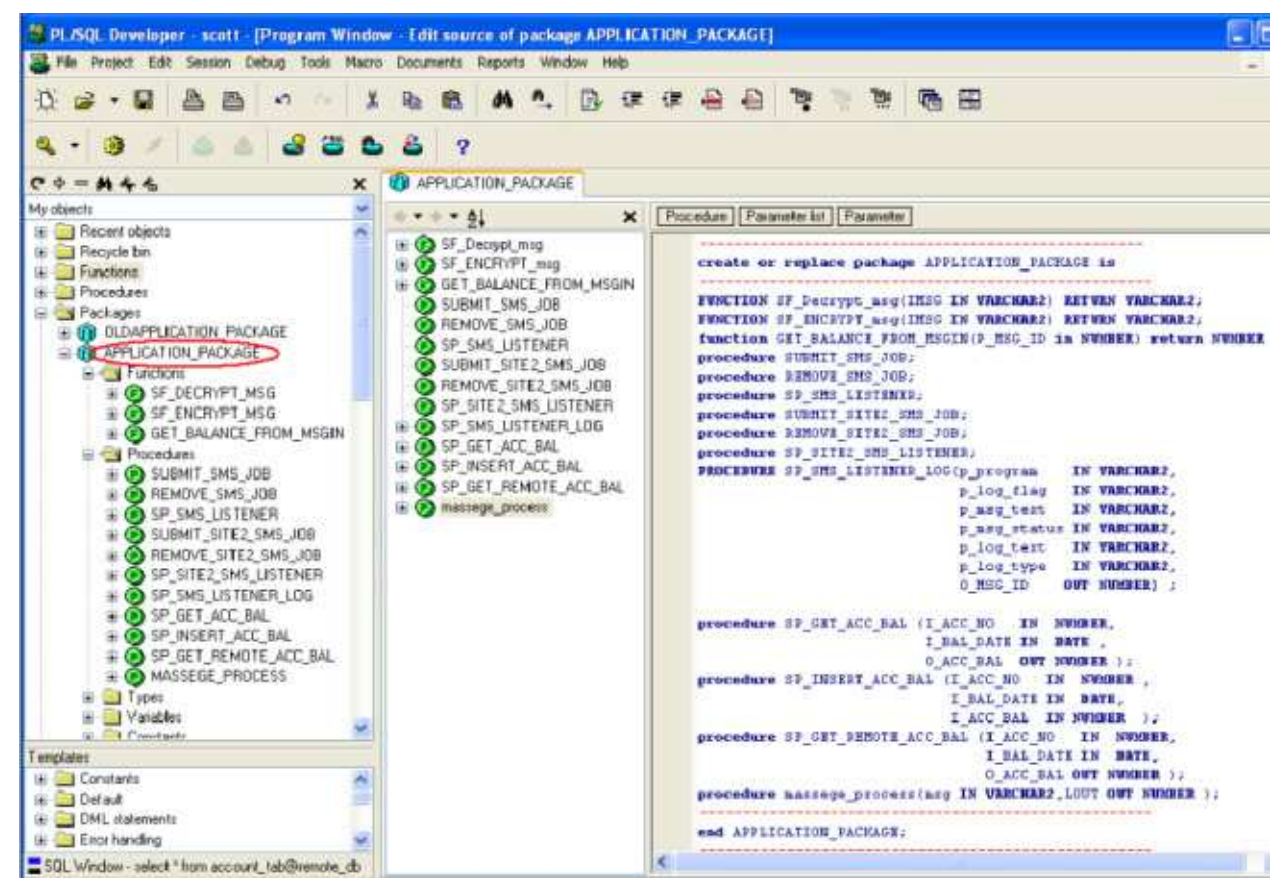

Figure 3. Application package components

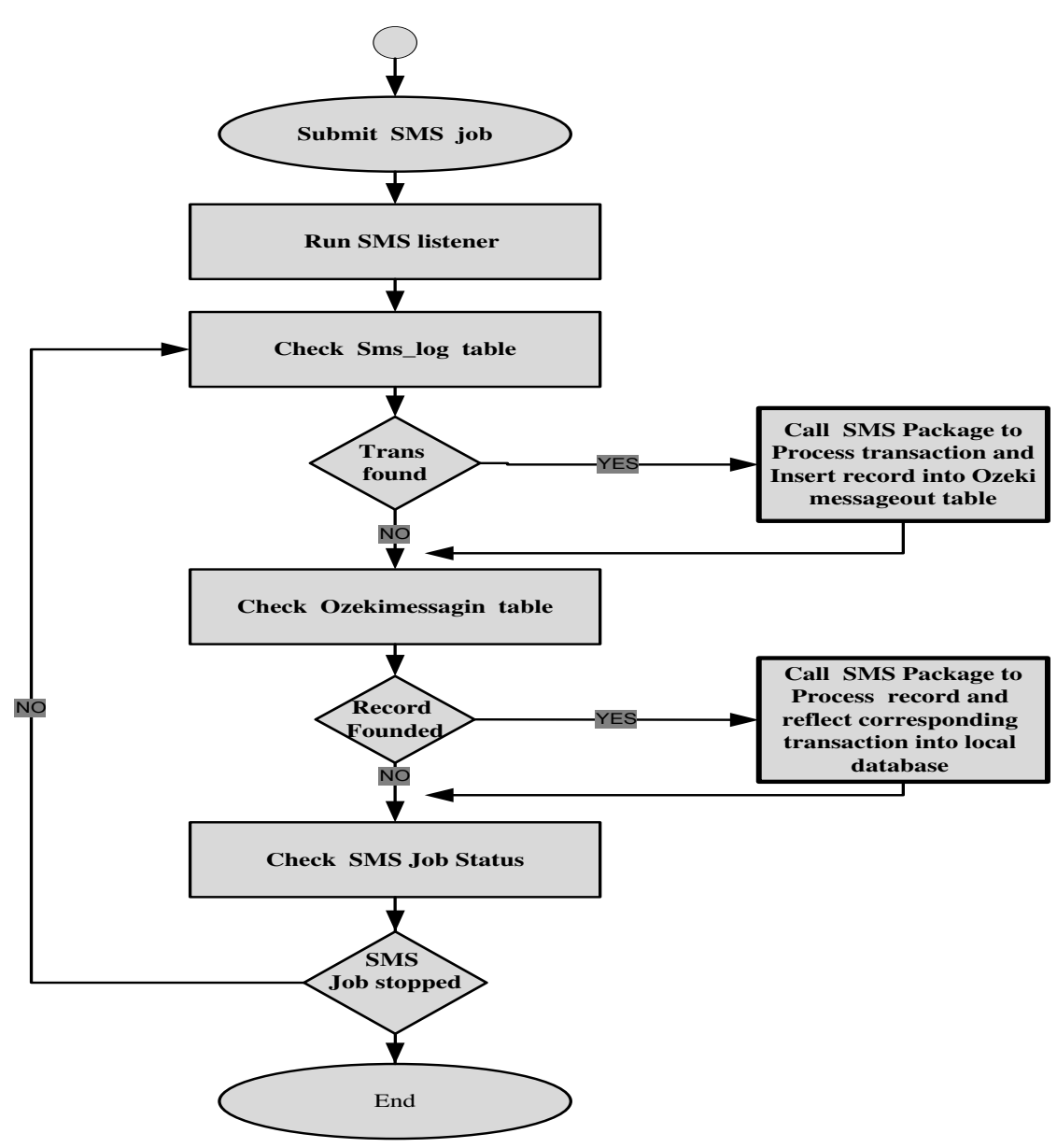

Figure 4. Flowchart of the programmed listener

## VI. SCENARIOS IMPLEMENTATIONS AND ANALYSIS

In order to test and measure the reliability of our model, we present four case-scenarios; these scenarios are related to the main functionalities that our model should do. We use two oracle database schemas; each schema, is related to a particular database user, one for Scott user and the other for the system user. The Scott user is the local database user within our model; and, the system user is the remote database user.

The four scenarios are:

(1) How to recover from a database-link disruption;

(2) Alerting the key person regards suspicious Transactions and/or situation;

(3) Alerting the Database Administrator (DBA) Regards invalid database objects;

(4) Submitting query remotely in other databases.

We carried out all of the four scenarios through our local database (the Scott database user) in order to modify the account balance table, which exists in the remote database (system database). The normal modification occurs through a predefined database link between the local and remote databases.

A. Algorithm Scenario One;

This scenario shows how we can recover from a database-link disruption when we trying to modify specific account balance through database link.

The following algorithm describes our current scenario:

Algorithm Scenario One;

Input: Account number, transaction amount,
Key-person mobile phone number,
Site2 mobile number;
Output: SMS message to alert key-person, notification message that account balance modification has been submitted remotely successfully.
Begin
Step 1: Make sure that the database link is Available and make a modification on a particular account balance of a Particular account number on the Remote database;
Step 2: Make sure that the modification Occurred;
Step 3: Disrupt the database link without Activating our model (SMS-Channel and SMS-JOB are OFF), and make the same modification in step 1 and then Make sure that the modification has not been submitted;
Step 4: Keep the database link disrupted and Activate our model (SMS-Channel and SMS-JOB are ON) and make the same Modification in step 1 and then make Sure that the modification has been submitted successfully;
End;

We carried out this scenario twice and it supported the account balance modification while the database link is disrupted. As a result, our model is reliable in terms of this test scenario.

B. Algorithm Scenario Two;

This scenario shows how we can employ SMS in our applications to alert the key person regards suspicious Transactions and/or situation.

The following algorithm describes the second scenario:

Algorithm Scenario Two;
Input :Account number, transaction amount, Key-person mobile phone number;
Output: Alerting SMS message to key person Regards Suspicious Transactions and/or Situation;
Begin
Step 1: Activate the SMS-Channel (Become ON) and the SMS-JOB (Submitted), and then make a Particular modification on the Account balance value regards a Particular account number on our Local database. The transaction should exceed the allowable Account value which is pre-defined in the business rules;
Step 2: Make sure that the modification has been occurred and then check the SMS-log-file table to find the related message of the modification.
Step3: Finally, check if the message has been inserted into the OZEKIMESSAGEOUT table in the Local database as an outgoing Alerting massage. At the end, this Message should be received by the key-person in his mobile phone;
End;

We carried our each step of this scenario twice to measure the test-retest reliability; were each scenario step should be identical in both times.

C. Algorithm Scenario Three;

This scenario shows how we can alert the Database Administrator (DBA) regards invalid database objects. The following algorithm describes the third scenario:

Algorithm Scenario Three;
Input : DBA mobile phone number;
Output: Alerting SMS message to DBA Regards invalid database object;
Begin
Step 1: Activate the SMS-Channel (Become ON) and the SMS-JOB (Submitted), and then enforce a Particular database Object to become invalid;
Step 2: Make sure that the message has been inserted into the SMS-log-file Table;
Step 3: Check if the message has been Inserted into the OZEKIMESSAGEOUT table in the Local database as an outgoing Alerting massage. At the end, this Message should be received by the DBA in his mobile phone.
End;

At the end of our scenario, the outgoing message in the ZEKIMESSAGEOUT table should be same as the message that shall get to a correspondent DBA mobile phone. Then the DBA shall modify and recompile the procedure, and this is going to save time and efforts. Again, we carried our each step of this scenario twice to measure the test-retest reliability; were each scenario step should be identical in both times.

D. Algorithm Scenario Four;

This scenario shows how we can submit query remotely in other databases. The following algorithm describes our current scenario:

Algorithm Scenario Four;
Input : Account number ;
Output: Account balance;
Begin
Step 1: Make sure that the database link is Available and then submit a Particular query on the remote Database to get balance for a Particular account number;
Step 2: Disrupt the database link within our Model (SMS-Channel and SMS-JOB Are OFF), and submit the same query in step 1 and then make sure that the Query has not been fetched;
Step 3: Keep the database link disrupted and Activate our model (SMS-Channel and SMS-JOB are ON) and submit the same Query in step 1, and make Sure that the Data has been retrieved successfully in the local database in Order to continue;
End;

We carried our each step of this scenario twice to measure the test-retest reliability; were each scenario step should be identical in both times.

Figure 5 shows four sequential steps regards scenario four.

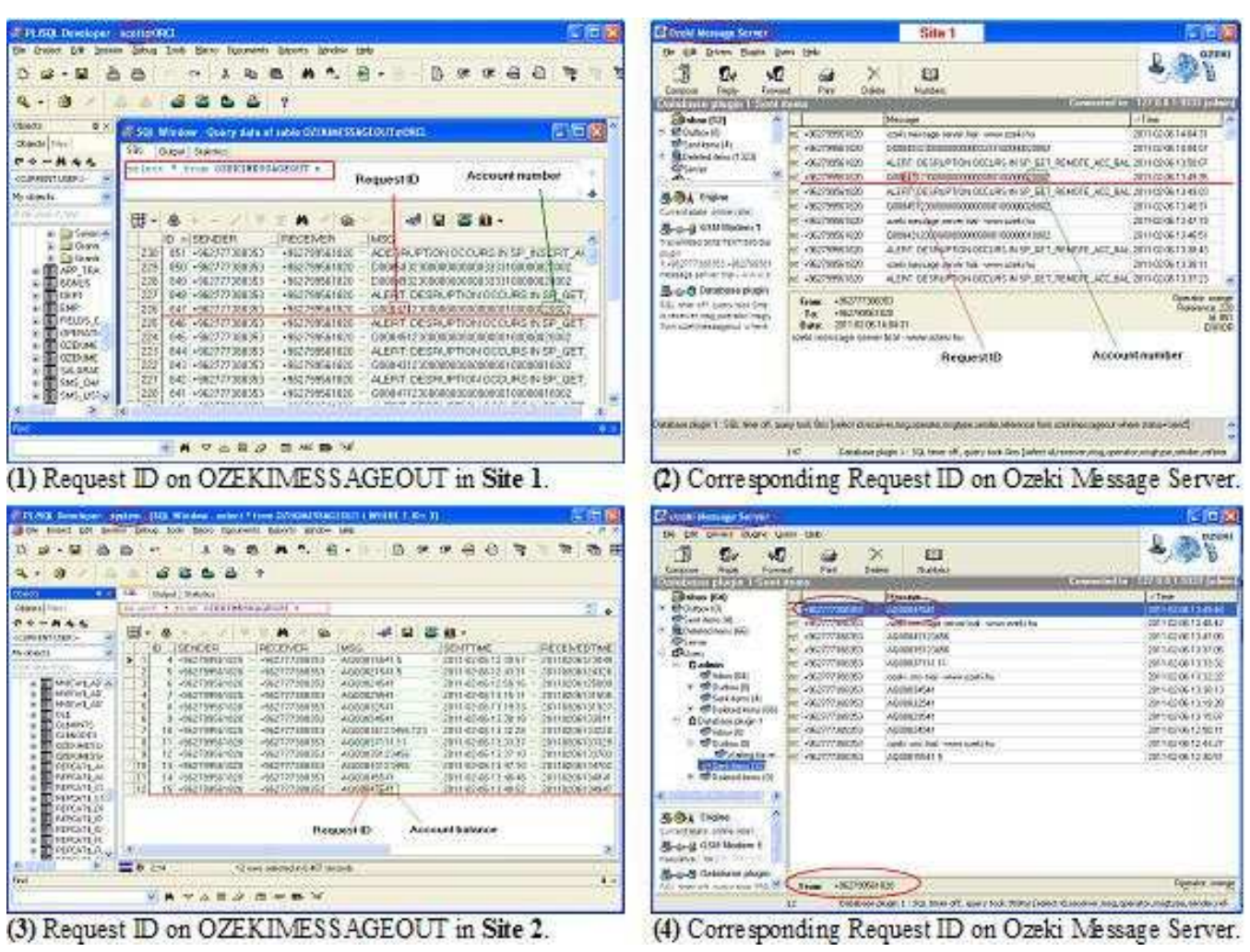

(1) Request ID on OZEKIMESSAGEOUT in Site 1. (2) Corresponding Request ID on Ozeki Message Server. (3) Request ID on OZEKIMESSAGEOUT in Site 2. (4) Corresponding Request ID on Ozeki Message Server.

Figure 5. Snapshots taken regards scenario four.

## VII. CONCLUDING REMARKS AND FUTURE WORK

In this research, we made a first step towards developing a new model of business continuity using SMS and other information system concepts such as organized databases, SMS listeners, and the SMS Ozeki server and others. We deploy our model into a test business domain, which is a computer laboratory in the computer department/ Amman Arab University. We carry out four mentioned scenarios; as a result to the experimental work, we conclude that our model is reliable and supports business continuity.

As a future work, SMS technology is still one of the most hot research topics; this importance is a result for the critical role that this technology plays in communication and the transmission of data and commands. in this regards, and since we are in the area of banking economy and competition; banking organizations are requiring robust and dynamic ATM applications through which customers can carry out their money transactions using their mobiles. As a result, within ATM machines, we need to provide an alternative standby channel that is based on the SMS technology in case any disruption occurs.

## REFERENCES

[1] Kogan Page Limited, THE SECURE ONLINE BUSINESS HANDBOOK e-commerce, IT functionality & business Continuity, 2004, second edition ISBN 07494 42212
[2] SMS Pal, Inc, Text Messaging Basics for Business, Version: 2008.05.01, 2008. White Paper
[3] Overland Storage, 2010, a Practical Guide to Business Continuity. White paper.
[4] The Benefits of Business Process Automation, through Site: http://miketurco.com/benefits business-process-Automation-10256
[5] James Kadirire, The short message service (SMS) for Schools/conferences, 2009
[6] Ibrahim A.S.Muhamadi, "Auto Notification Service for The Student Record Retrieval System Using Short Message Service (SMS)"; International Journal of Computer Science and Network Security, VOL.9, No.8, August 2009
[7] Ozeki Informatics Ltd, Ozeki Message Server 6 Product Guide, 2006
[8] Michael Gallagher, 2003, Business Continuity Management: How to protect your company from Danger. ISBN: 0273663518, Prentice Hall
[9] The Government of the Hong Kong Special Administrative Region, SHORT MESSAGE SERVICE SECURITY, February 2008.
[10] Intermec Technologies Corporation, Eliminating paperwork Is More Than Just Efficient, 2008 whitepaper.
[11] Edy Jordan, INTERFACING SMS AND DATABASE SYSTEMS: ASOFTWARE ENGINEERING APPROACH, 2004
[12] Twenty First Century Communications, High-Volume Inbound IVR – Critical for Business Continuity , White Paper, through site: www.tfcci.com
[13] Dynmark International org, sending out an SMS: Texting in an emergency, September 2010, white paper.
[14] Andreas Rosendahl, J. Felix Hampe, and Goetz Botterweck, Mobile Home Automation, Merging Mobile Value Added Services and Home Automation Technologies, Proceedings, Sixth International Conference on Mobile Business, 8–11 July 2007, IEEE Computer Society, ISBN 0-7695-2803-1".